\documentstyle[epsfig]{mn}
\newcommand{\overdot}[1]{\stackrel{.}{#1}}
\newcommand{\MJ}{$M_J$}
\newcommand{\delete}[1]{}
\newcommand{\add}[1]{#1}

\begin{document}

\title[Reversing type II migration]{Reversing type II migration: resonance trapping of a lighter giant protoplanet}
\author[F. Masset \& M. Snellgrove]{F. Masset \thanks{send offprint
requests to: {\tt F.S.Masset@qmw.ac.uk}} 
and M. Snellgrove\\
 Astronomy Unit,  School of Mathematical Sciences, 
 Queen Mary \& Westfield College, Mile End
 Road, London E1 4NS, England\\
}

 \date{Received *****; in original form ******}


\maketitle

\begin{abstract}
We present a mechanism related to the migration of giant protoplanets embedded
in a protoplanetary disc whereby a giant protoplanet is caught up, before
having migrated all the way to the central star, by a lighter outer giant protoplanet.
This outer protoplanet \add{may} get
captured into the 2:3 resonance with the more massive one, \delete{and}\add{in which case}
the gaps that the two planets open in the disc overlap.
Two effects arise, namely a \add{squared} mass weighted torque imbalance and
an increased mass flow through the overlapping gaps from the outer disc
to the inner disc, which both play in favour of an outwards migration. Indeed under
the conditions presented here, which describe the evolution of
a pair of protoplanets respectively Jupiter and Saturn sized,
the migration is reversed, while the planets semi-major axis ratio is constant
and the eccentricities are confined to small values by the disc material.
The long-term behaviour of the system is briefly
discussed, and could account for the high eccentricities observed for the extrasolar
planets with semi-major axis $a>0.2$~AU.
\end{abstract}


\section{Introduction}
In the past few years a number of extrasolar giant planets have been
discovered around nearby solar--type stars. These objects masses range from
$0.17$~\MJ~to $11$~\MJ~(where \MJ~is Jupiter's mass) and their orbital
semi-major axis range from $0.038$~AU to $3.3$~AU (Marcy, Cochran \& Mayor,
1999). 
Although many uncertainties remain about planet formation, it is now
commonly accepted that planets have formed in and from protoplanetary discs.
Necessarily,
there must be some time interval over which a giant planet and 
the surrounding gaseous disc material coexist.
The planet and the disc exchange angular momentum through tidal 
interactions which generally make the planet lose angular momentum
This mechanism is called
migration. It can roughly be divided
in two regimes:
\begin{itemize}
\item If the planet mass is small enough, the disc response
is linear. The migration
rate, in that regime, is proportional to the planet \add{and disc}
mass\add{es}, is independent
of the viscosity and weakly dependent of the disc surface density
and temperature profiles. This is the so-called type~I migration (Ward, 1997).
\item When the protoplanet mass is above a certain threshold, 
the torques acting locally on the surrounding disc material open a gap (Papaloizou \& Lin, 1984),
whose width and depth are controlled by the balance between the tidal
torques, which tend to open the gap, and the viscous torques which tend
to close it.
The disc response is \delete{markedly} \add{significantly} non-linear, and
most of the
protoplanet Lindblad resonances fall in the gap and therefore
cannot contribute to the planet-disc angular momentum exchange. The 
migration rate slows down dramatically compared to type~I migration.
Furthermore, the tidal truncation splits the disc in\add{to} two parts and
the
planet is locked \delete{in} \add{to} the disc viscous evolution (Nelson
et al.
2000).
This is the type~II migration, \add{which describes the orbital evolution of giant
protoplanets.} 
\end{itemize}

\delete{Type~II migration occurs
on the disc viscous time-scale, about $ 10^5$~yr at 5~AU. 
This time-scale is still one or
two orders of magnitude shorter than the disc estimated lifetime.
Unless the planet has formed much further out in the disc, in which
case one has to face time-scale problems again, or unless the disc
is cleared by some process before the planet has migrated all the
way to the central star,
there is clearly
a conflict between type~II migration theoretical predictions
and observational facts. Indeed, roughly speaking only half
of the detected extrasolar planets are ``hot Jupiters'', which
orbit very close to the central star ($a<0.2$~AU).

There is another problem with the type~II migration of one giant
protoplanet. All the extrasolar planets
with a semi-major axis larger than $0.2$~AU have large eccentricities
($e>0.16$, except 47~UMa for which $e=0.096$, which is already twice as large as Jupiter's
eccentricity). Now the migration of a giant protoplanet
embedded in a disc occurs on a nearly
circular orbit, since the eccentricity is damped by the coorbital material (Ward, 1988).}

In this letter we consider the coupled evolution of
a system of giant protoplanets consisting of two non-accreting
cores with  masses $1$~\MJ~and
$0.29$~\MJ, which we are going to call from now on respectively
``Jupiter'' and ``Saturn''. 
Attempts have already been made to describe the behaviour of a system
of planets embedded in a disc. Melita \& Woolfson (1996) and
Haghighipour (1999) considered an embedded Jupiter
and Saturn system orbiting a solar mass star, and showed how resonance
trapping would affect their evolution. However 
the dissipative force in these works  was
due to the dynamical friction with a uniform density interplanetary medium,
hence type~II migration effects were not taken into account. 
 Resonance trapping of planetesimals
by a fixed orbit Jupiter sized protoplanet has also been investigated by Beaug\'e et al. (1994),
and shown to be able to build up a single planetary core with orbital 
characteristics close to Saturn's ones.
Kley (2000)
studied the orbital evolution of two maximally accreting giant cores embedded in a
minimal mass protosolar disc, and 
showed how the migration of the inner core could be halted
by the presence of the outer one, and how the eccentricity of the inner
core is pumped up by the outer one.\delete{and hence could account for observations
of high eccentricity extrasolar planets.}

\section{Results}

\subsection{Numerical codes description}
In order to investigate the long-term behaviour of the embedded Jupiter and
Saturn system, we have used two independent hydrocodes, which have
been described elsewhere in full detail (Nelson et al. 2000). These two
codes are fixed Eulerian grid based codes, one of them is NIRVANA (Ziegler \& Yorke, 1997) and
the other one has been written by one of us (FM). Both have been endowed 
with the fast advection FARGO algorithm (Masset, 2000), \add{and can run
either with
this algorithm or with a standard advection algorithm}. They gave very similar
results. They consist of a
pure N-body kernel based on either a fourth (NIRVANA) or fifth order adaptative timestep
Runge-Kutta solver 
(sufficient for the short time-scales involved in this
dissipative problem) embedded in a hydrocode which provides a tidal interaction
with a 2D non self-gravitating gaseous disc. The simulations are performed in the
non-inertial non-rotating frame centered on the primary. The grid outer
boundary does not allow inflow nor outflow and is chosen sufficiently far 
from the planets in order for the spiral density waves that they launch to be damped
before they reach it,
while the grid inner boundary only allows 
outflow (inwards), so that the disc material can be accreted on to the primary.
Failing to do so may lead to overestimate the inner disc density and artificially
favours an outwards migration. In the following our length unit is
$5.2$~AU, the mass unit is one solar mass,
and the time unit is the initial orbital period of Jupiter (the actual period may
vary as Jupiter migrates). The disc aspect ratio $H/R$ is uniform and constant.
In the run presented here the grid resolution adopted is
$N_r=122$ and $N_\theta=300$ with a geometric spacing of the interzone radii
---~such that all the zones are ``as square as possible'', i.e.
$N_r\log(1+2\pi/N_\theta)=\log(R_{max}/R_{min})$~---.
The grid outer boundary is at $R_{max}=5$ and its inner boundary is
at $R_{min}=0.4$. 
The geometric spacing
is the most natural one since the disc thickness
scales as $r$. On the other hand, a constant spacing
leads to an oversampling of the outer disc and an undersampling of the
inner one, and \delete{hence}\add{therefore} is likely to favour an inwards migration.

\subsection{Initial setup}
The cores we consider are embedded in
a gaseous minimal mass protosolar nebula around a unit mass central object, 
and we assume they start their
evolution with semi-major axis $a_j=1$ for Jupiter and $a_s=2$ for
Saturn. The disc surface density is 
uniform and corresponds to  two Jupiter masses 
inside Jupiter's orbit. The effective viscosity $\nu$, the nature
of which remains unclear and is usually thought to be due to turbulence generated
by MHD instabilities (Balbus \& Hawley 1991), is assumed to be uniform through
the disc and corresponds to a value of $\alpha \simeq 6\cdot 10^{-3}$ in the vicinity
of Jupiter's orbit. The disc aspect ratio is  $H/r=0.04$.

The mass of Jupiter is sufficient to open a \delete{clearly
marked}\add{deep} gap
and hence it settles in a type II migration (Nelson et al. 2000), whereas
Saturn is unable to fully empty its coorbital region because:
(i) its mass is smaller;
(ii) The planet is in a regime \delete{called}\add{known as} the 
inertial limit (Ward \& Hourigan, 1989\delete{; Nelson, 2000}) where
the inwards migration speed is so high that it makes the planet pass through what
would be the gap inner
edge before it had time to actually open it.

Therefore Saturn \delete{has almost no}\add{does not clear a deep} gap
\delete{in the early stages}\add{initially}, 
and
its migration rate is typical of  type~I migration, 
since all
its  Lindblad resonances \add{can still} contribute to
\delete{exchange}\add{the} angular momentum
\add{exchange}
with the disc.

\subsection{Run results}
\begin{figure}
  \begin{center}
    \psfig{file=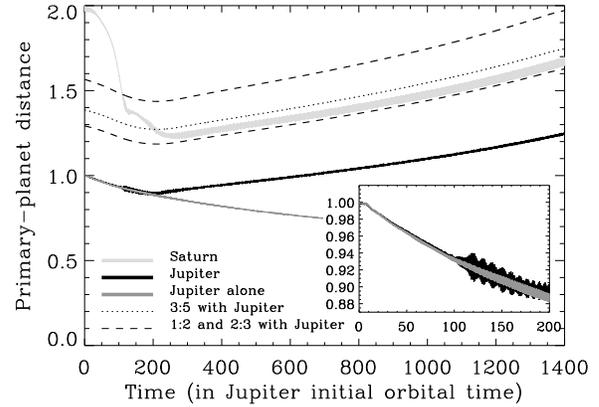,width=\columnwidth}
    \caption{Primary--planet distances as a function
of time.
The outer dashed curve represents the nominal position of the 1:2 resonance with Jupiter, while
the inner dashed curve is the nominal position of the 2:3 resonance. The zoomed plot enables one
to closely compare Jupiter's orbital evolution against a test run without Saturn.}
    \label{fig:mig}
  \end{center}
\end{figure}

\begin{figure}
  \begin{center}
    \psfig{file=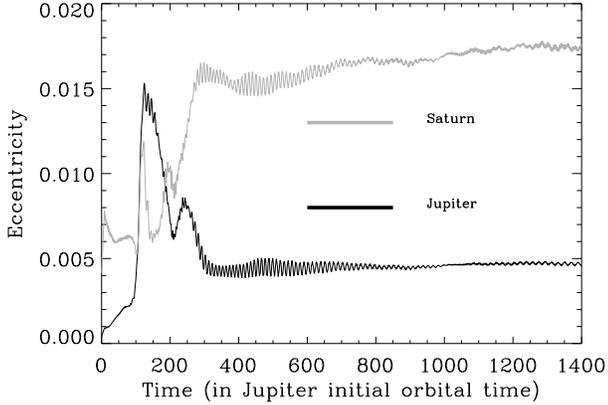,width=\columnwidth}
    \caption{In this figure we see the planets eccentricities as a function
of time. They simultaneously increase as Saturn passes through the 1:2 and
3:5 resonance\add{s} with Jupiter. Once Saturn is trapped in\add{to the} 
2:3 resonance
with Jupiter, both eccentricities
settle at a roughly constant level, which results of a balance between the migration rate
which pumps them up
and the eccentricity damping by the disc coorbital material. \delete{(through the pile-up of the 
outer fast eccentric
Lindblad resonances and inner slow eccentric Lindblad resonances).}
}
    \label{fig:ecc}
  \end{center}
\end{figure}

We present in fig.~\ref{fig:mig} the central star--planet distance curves
as a function of time. We see how initially Jupiter migrates as if it
was the only planet in the disc (see test run). In the meantime, Saturn
starts \add{a} much faster \delete{a} migration (\delete{after some delay
which corresponds to the disc response 
time-scale}\add{the obvious initial acceleration of its migration will be discussed 
elsewhere}),
and reaches the 1:2 resonance with
Jupiter at time 
$t\simeq 110$.
The eccentricities at that time are
small (see fig.~\ref{fig:ecc}), and in particular Saturn's eccentricity is much smaller than 
the eccentricity threshold below which the capture into resonance is certain
if the ``adiabatic'' condition on the migration rate is satisfied (Malhotra 1993):
$|\!\!\overdot a_s\!\!|/(a_s\Omega_s) \ll 0.5j(j+1)\mu_Je_s$ for the $j$:$j+1$ resonance, where $\mu_J$ is the mass
ratio of Jupiter to the central object, \add{and where $e_s$ is Saturn's eccentricity}. This condition is not satisfied when Saturn reaches the
1:2 resonance, and it passes through.

The planets then \delete{have increased}\add{obtain higher}
eccentricities, and Saturn's migration rate is reduced.
\delete{The reason for \delete{that}\add{this} will appear later, and has
to
\delete{see}\add{do}
with an
increased inwards mass flow.} 
Saturn's eccentricity increases again rapidly as it passes through the 3:5 resonance with
Jupiter at 
$t\simeq 220$. 
Eventually the adiabatic condition on the migration rate is satisfied for the 2:3 resonance
and Saturn's
eccentricity is still below the corresponding critical threshold, so it
gets trapped in\add{to the} 2:3 resonance
with Jupiter (both $e$ and $e'$ resonances, since the two critical angles $\phi=
3\lambda_s-2\lambda_j-\tilde\omega_s$ and $\phi'=3\lambda_s-2\lambda_j-\tilde\omega_j$
librate, where $\lambda$ is the mean longitude
and $\tilde\omega$ the longitude of perihelion). At that time both planets
steadily migrate outwards.

%

\subsection{Interpretation}
\begin{figure}
  \begin{center}
    \psfig{file=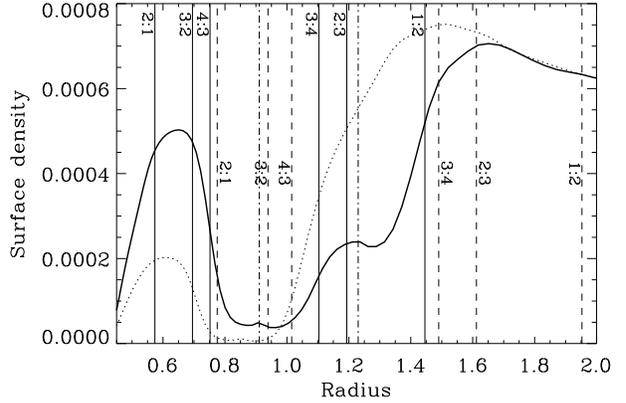,width=\columnwidth}
    \caption{Azimuthally averaged surface density as a function of radius, for the two
planet run (solid curve) and for the test run with Jupiter only (dotted curve), at time
$t=286$~orbits. The solid
vertical lines show Jupiter's circular Lindblad resonances, and the dashed lines 
Saturn's circular Lindblad resonances. The dot-dashed lines at $r=0.91$ and $r=1.23$ show respectively
the positions of Jupiter (in the two planet run) and Saturn. As can be seen also in fig.~\ref{fig:mig},
the Jupiter to Saturn orbital ratio is slightly larger than $3/2$. This is due to the fast precession
of the perihelions.}
    \label{fig:lr}
  \end{center}
\end{figure}

\delete{The behaviour described above has a twofold origin.}
\delete{\bf A mass weighted torque imbalance:}\add{We define
the system of interest as the system composed of the two planets. This}
 resonance locked system \delete{composed
of the two planets} interacts with the inner disc through torques proportional to
$M_J^2$, at Jupiter's inner Lindblad resonances (ILR), 
whereas it interacts with the outer disc through torques proportional
to $M_S^2$ at Saturn's outer Lindblad resonances (OLR), as indicated on fig.~\ref{fig:lr}.
\delete{Now}\add{It can be seen that} Saturn's ILR fall in Jupiter's gap
 and
Jupiter's
OLR fall in Saturn's gap so their effect is weakened compared to the situation
where Jupiter is alone. \delete{Now}\add{As} $M_J^2/M_S^2\sim 10$,
\delete{hence}
the torque imbalance does not favour an inwards migration as strongly as in a one planet case,
and \delete{might}\add{may} even lead to a positive \delete{net}\add{differential
Lindblad} torque on the two planet system.
\add{Actually one can estimate what the maximum mass ratio of the outer
planet to the inner one should be
to get a migration reversal, if one neglects the Inner Lindblad torque  on the outer planet and the
Outer Lindblad torque on the inner planet. The Inner Lindblad torque on
the inner planet reads as:
\begin{equation}
\label{eq:tin}
  T_{ILR}=C_{ILR}\mu_J^2\Sigma_0a_J^2(a_J\Omega_J)^2h'^{-3}
\end{equation}
where $C_{ILR}$ is a dimensionless coefficient which is a sizable fraction
of unity (Ward, 1997), and where $h'$ is the disc aspect ratio.
There is a similar formula for the Outer Lindblad torque on the outer
planet (obtained by substituting
the~$ILR$ and~$J$ indices in Eq.~(\ref{eq:tin}) respectively with~$OLR$
and~$S$). The resulting torque imbalance
will be positive if: $T_{ILR}>T_{OLR}$, which reads here as:
\begin{equation}
  \label{eq:raplim}
  \frac{\mu_S}{\mu_J}<\left(\frac{C_{ILR}}{C_{OLR}}\right)^{1/2}\left(\frac 23\right)^{1/3}
\end{equation}
If we assume that $C_{ILR}=C_{OLR}$ then we get: $\mu_S/\mu_J < 0.87$,
whereas if we make the conservative
assumption that $C_{ILR}=\frac 12 C_{OLR}$, we have: $\mu_S/\mu_J < 0.62$.
This threshold is much bigger
than the actual ratio, therefore if the common gap is deep enough to shut off Jupiter's OLR torques 
(and Saturn's ILR torques)
then the net Lindblad torque on the two planet system is positive.}
\delete{{\bf An increased mass flow} from the outer disc to the inner disc through the common
gap. Indeed, one can see in fig.~\ref{fig:flow} that the mass flux crossing Jupiter's orbit
is about four times higher
in the two planet run with respect to the test case with Jupiter only.}
\add{As the two planet system proceeds outwards in the disc, it does not act on the gas
as a snow-plough, but rather it allows the material from the outer disc to
travel across
the common gap and eventually feed the inner disc. We can find the gap
``permeability''
condition by requiring that the rate of angular momentum change of
the ring
of material lying immediately outside Saturn's gap that is 
 required to expand accordingly to Saturn's orbit  (snow-plough
effect) is greater than the torque available from Saturn (at most the sum of its outer
Lindblad torques, in which case we need to assume that the waves excited at their OLR
are damped locally). In our case, this turns out not to be the case and most  of the
outer disc material flows through the common gap to the inner disc.
We find that in all our runs it is possible to
check that the rate of mass
flow through the common gap (see fig.~\ref{fig:flow}) can be expressed as~:
\begin{equation}
  \label{eq:flux}
  \overdot M\simeq 3\pi\nu\Sigma_0+2\pi r_s\overdot r_s\Sigma_0
\end{equation}
with a reasonable precision ($10-20$~\%). Furthermore we have performed many ``restart
\label{sec:restart}
runs'' which consist in restarting a run once Jupiter and Saturn are
locked into resonance, and then 
by varying one parameter at one time, e.g. the viscosity or the aspect ratio (which
changes the Lindblad torques and therefore the migration rate). The mass flux through
the gap rapidly switches (in a few tens of orbits) to a new value after
the
restart,
so that Eq.~(\ref{eq:flux}) remains fulfilled. 

From the considerations above we can conclude that the presence of Saturn unlocks
Jupiter from the disc evolution~: the two planet system evolution (outwards) and the
disc viscous evolution (inwards) are basically decoupled.
}
This \add{decoupling and the corresponding} mass flow \add{through the 
common gap} has two consequences~:
\begin{figure}
  \begin{center}
    \psfig{file=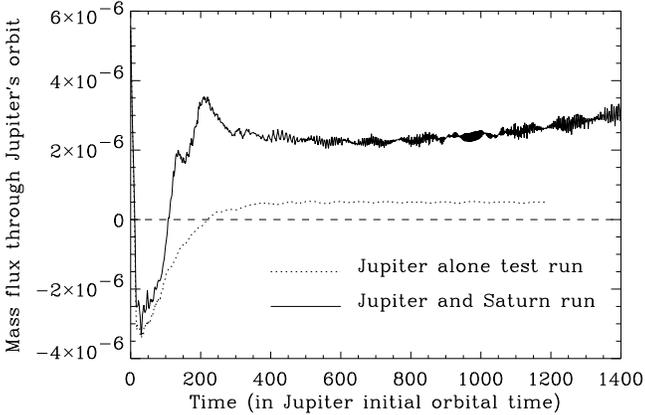,width=\columnwidth}
    \caption{Mass flux crossing Jupiter's orbit (in mass units per orbital
time), \add{where} positive \add{is} for
an inwards flow. This quantity can be estimated from the total amount of mass located outside
Jupiter's orbit (more precisely outside a circle having a radius equal to Jupiter's semi-major axis,
in order to smooth out the short period variations linked to the eccentricity), 
since the outer boundary is closed ($v_{\rm rad} \equiv 0$). The negative
value at the early stages is due to a relatively fast inwards migration; it reverses
for both runs, even for the Jupiter only test run (where the inner disc is rapidly depleted). Note
that the mass flux is reversed {\em before} the migration reverses, when Saturn
passes through the 1:2 resonance, \add{which more or less corresponds to
the time at which the coorbital
regions of both planets merge}.}
    \label{fig:flow}
  \end{center}
\end{figure}
\begin{itemize}
\item A \add{re}filling of the inner disc, which is \delete{normally} too
depleted for the torques
at Jupiter's ILR to have any sizable effect in the one planet case (the inner
disc is accreted on the primary on its short viscous time-scale
and maintaining its surface density at not too low a value implies a \delete{high 
flow rate of material} \add{permanent flow of material}
from the outer disc \add{to the inner}).
\item  The angular momentum lost by the material which flows from the outer disc to the inner
one is gained by the planets.
\add{The exchange of angular momentum between a planet and a gas fluid
element occurs
during a ``close encounter'' between these two, the one planet version of which
corresponds to the angular momentum exchange at each end of a horseshoe
orbit of the fluid element. The resulting
torque is the so-called coorbital
corotation torque (Goldreich \& Tremaine 1979, Ward 1991 and 1992). 
To the
best of our knowledge, an analytical evaluation of the corotation torque in the case of
a non-vanishing net mass flow through the orbit (either due to a viscous
accretion on to the primary or to a radial migration or both) has not been performed
yet. Obviously even the one planet case deserves a large amount of work on this
specific topic, therefore the estimate of the corotation torque in this two planet
problem is far beyond the scope of this paper. We will just comment that
the corotation torque in our case 
might not be negligible compared to the differential
Lindblad torque at some stage.}
\end{itemize}

\section{Discussions}

\add{We have performed a series of restart runs (see
section~\ref{sec:restart}) in order
to check for a variety of behaviours.

\subsection{Differential Lindblad torque sign}
The one sided Lindblad torque has been shown to be proportional to $h'^{-3}$ (Ward 1997). We
have performed two restart runs ($h'=0.04\rightarrow 0.03$ and $h'=0.04\rightarrow 0.05$)
in order to check that the migration rate variation is consistent with this dependence.
This is indeed the case. We note in passing that the migration rate varies
as
$h'^{-3}$, and not as $h'^{-2}$ as it would be the case in a one planet
problem, since
the Outer/Inner Lindblad torque asymmetry does not vanish as the disk thickness tends
to zero (the OLRs would pile-up at Saturn's orbit, whereas the ILRs would pile-up
at Jupiter's orbit). These results confirm that the behaviour we observe
occurs mainly due to the
differential Lindblad torque and shows as well that this latter quantity
is positive, as expected from Eq.~(\ref{eq:raplim}).}

\subsection{$\alpha$-viscosity vs. uniform viscosity}
So far we have only considered a uniform viscosity. Switching to
a uniform-$\alpha$ viscosity of the form $\nu=\alpha c_sH$ \delete{would} makes $\nu$ scale here
as $r^{1/2}$, so the viscosity
at the outer edge of the common gap \delete{would be}\add{is} higher, whereas it \add{is}
\delete{would
be} smaller in the inner disc. This \delete{would} has \delete{two}\add{the
following} effect, which \delete{would both} plays in
favour of enhancing the migration reversal mechanism:
\delete{(i) The mass flow through the gap, which is thought to be linked to the
viscosity, is likely to increase as the viscous stress increases (since the tidal
truncation is weaker and more material is allowed to penetrate the gap on Saturn's side);}
the viscous time-scale of the inner disc \delete{would be}\add{is} higher and therefore
its surface density \delete{would} increases \delete{as well}\add{accordingly}, 
since the material brought through
the gap \delete{would} piles-up in the inner disc for a longer time before being
accreted on the primary. \add{This has been checked with a
restart
run}.

\subsection{Accretion on to the planets}
The cores considered above do not accrete gas from the
disc. One can wonder what would be the effects of accretion.
\add{We have performed a number of restart runs in order to investigate
the effect of
accretion on the mechanism presented here. We have only considered accretion on to
Jupiter, as it is likely that the accretion rate on Saturn can be
regarded as being negligible (i.e.
its mass doubling time is much longer than the timescale of the outwards migration,
see e.g. Pollack et al. 1996). The prescription we used to model accretion
on to Jupiter
consists in removing a proportion of the material which lies in the inner
Roche lobe
(i.e. a sphere with a radius of half the Hill radius). The amount which
is removed in one timestep is calculated from the half emptying time
of the inner Roche lobe
$\tau_{1/2}$. We have performed four different restart runs, corresponding to the
following values of $\tau_{1/2}$: $\tau_{1/2}=T_0$ (maximally accreting core, see Kley
1999),
$\tau_{1/2}=3T_0$, $\tau_{1/2}=10T_0$ and $\tau_{1/2}=30T_0$, where $T_0=2\pi/\Omega_J$
is Jupiter's orbital time. In each of these cases, turning on accretion
had no 
impact on the system migration rate, at least in the early stages: in the first
case, the mass doubling time  for Jupiter is relatively short, and when
Jupiter's mass is
significantly larger than its initial mass some additional effects, which will be
presented in much greater detail elsewhere, affect the migration rate
which then
differ from the non-accreting case.

}

\subsection{\delete{Grid resolution and }Smoothing}
\delete{The mechanism presented above relies on an increased mass flow through
the overlapping gaps.  No analytical theory, at the present date, is able
to predict the flux of mass going from the outer disc to the inner disc, and
one can wonder whether the observed flux is not linked to numerical aspects,
 in particular the finite zone size. To answer this question,
we performed a test run with twice as high a resolution as the
geometric radial spacing run, i.e. with $N_r=244$
and $N_\theta=600$, all the other parameters being unchanged. No difference
has been found on the mass flow through the gap. }

The smoothing parameter of the potential can have a dramatic
impact on \delete{the initial} Saturn's \add{initial} migration
rate. \delete{Indeed} This rate is controlled by a subtle balance between
outer disc
and inner disc torques.
In the case of Saturn, all the Lindblad resonances
play a role, since there is no gap.
Many
prescriptions for the smoothing are unable to give trustworthy results for the balance
between the outer and inner torques since, depending on the prescription, these two
quantities are affected in a different way.
On the other hand Jupiter's migration
rate is much more robust, since the presence of the gap prevents high-$m$
Lindblad resonances \delete{to} play\add{ing} a role in the migration,
which is \add{therefore} controlled
only by remote, low $m$ resonances and \delete{therefore}\add{thus}
almost insensitive to the smoothing parameter. For this reason we
have adopted an approach which \delete{consists in}\add{involves} choosing
a smoothing
prescription  which endows Saturn with a migration velocity of the order
of magnitude of the linear analytical predictions (type~I migration),
which is needed to give correct results for the capture in\add{to} 
resonance.
Once Saturn is trapped into resonance with Jupiter, it is
dynamically slaved by \delete{this}\add{the} latter and the system
evolution is \add{only} very weakly
affected by the exact value of the outer disc torque exerted on Saturn.
We have found that using \add{either of} the \add{two} prescription\add{s} 
below satisfactorily preserves the analytical
torque imbalance on Saturn and therefore gives it a type~I migration rate:
\begin{itemize}
\item The potential of a planet acting on the disc is smoothed over the
length $\varepsilon=0.4R_H$ where $R_H$ is the Hill radius of the planet under
consideration, whereas the potential of the disc acting back on the planet
is smoothed over $\varepsilon'=\sqrt{H^2+d^2}$ where $H$ and $d$ are respectively
the local disc thickness and zone diagonal. Since $\varepsilon'\ne\varepsilon$
the action-reaction law is not fulfilled and the numerical biases which arise
favour an inwards migration, as can be easily checked.
\item The potential of a planet acting on the disc and the potential of the
disc acting on the same planet are smoothed over $\varepsilon = 0.4R_H$.
This prescription does fulfill the action-reaction law.
\end{itemize}
In \add{both} these two cases, as in any other which gives Saturn a type~I
migration rate,
including runs performed with a uniform radial spacing,
the migration gets reversed. The run presented here corresponds to the first
prescription.

\subsection{Impact of mass ratio and Long-term behaviour}
One can wonder \delete{on which}\add{about the size of the} interval of
``Saturn'''s mass \add{which causes} the migration \delete{can}\add{to} be
significantly 
slowed down or reversed. \delete{Indeed} If ``Saturn'' is not massive
enough it will not significantly
affect Jupiter's evolution (\add{the common ``gap'' will be too full on Saturn's side, 
and therefore Jupiter's OLR torques will not be shut off}), whereas if it is too massive,
the torque imbalance will \add{be negative again}.\delete{strongly play in favour of an inwards migration
and the
steepest tidal truncation at the common gap outer edge is likely to  reduce the inwards
mass flow.}
Work is in progress to accurately
determine which range of parameters leads to a migration reversal.
\add{It should be noted that the results presented here depend on the artificial initial conditions. We have
performed other runs in which Saturn is initially very close either to the $1:2$ or $3:5$ resonance, and
it turns out that neither of these resonances is able to struggle against
the strong Lindblad torques on
Saturn: no resonance angle can be found which provides a resonant torque on Saturn which counteracts 
the tide. Therefore a trapping into the $2:3$ resonance is the most likely outcome when the
system is still embedded in a massive disc, whatever the initial conditions: catching-up of
``Saturn'' or {\em in-situ} assembling from smaller, type~I migrating bodies.
}
\delete{, and
to properly understand the physics of the high inwards mass flow (funnel
effect due to the weak tidal truncation of the outer edge, or a possible
link with the stochastic behaviour of a test-particle which arises in the coorbital region of a pair
of resonantly orbiting massive planets). }

The long-term behaviour of the system  is twofold: 

\begin{itemize}

\item \delete{It should be
noted that the outwards migration  is a slowly accelerating process (the torques
scale as $a$ and the inwards mass flow increases as well for geometrical reasons)
which requires the gap outer edge to be able to follow the migration
rate $\overdot a/a$. Now the torque Saturn exerts on the outer disc
scales as $a$, whereas the angular momentum
content of the outer disc scales as $a^{5/2}$, so for a
sufficiently high value of $a$, the gap outer edge will not be able to follow
the global outwards motion, and a runaway inwards mass flow
can result. The most important consequence of this runaway flow is that Saturn
will leave the 2:3 resonance with Jupiter and will move further in the disc. The details and
consequences
of this mechanism will be presented in a forthcoming paper.}
\add{The system is locked in\add{to} resonance as long as~:
  \begin{itemize}
  \item The two-planet system can adjust its resonance angle in order to prevent the
planets being ``pushed'' towards each other by the
Lindblad torques exerted by the disk
on each of them. In all our runs we have never observed this behaviour. Now,
given the small eccentricities involved here, 
and given the fact that the adiabatic criterion threshold increases as $j(j+1)$, the
most probable outcome is that Saturn would then be captured in the
next order resonance,
that is to say $3:4$, and all the physics exposed in this paper would still be valid
(presence of a common gap, sharing 
of the coorbital material by the two planets, mass-weighted
torque imbalance, etc.)
\item The planets are not pulled apart by any other torques. Now we have mentioned
the possibly important role of the coorbital corotation torque in this problem,
which may be sufficient to move the planets apart at some stage, 
in which case we may ultimately get
a low eccentricity
double giant planet system when the disc disappears.
 This will be presented in greater detail elsewhere.
  \end{itemize}
}

\item
 If the planets happen to be \add{locked} in\add{to} resonance at the time
\add{that} 
the gas effects become
negligible,
then the system is likely to be unstable
(we  mentioned already that at
least two angles librate simultaneously, which strongly suggests
a possible chaotic behaviour; see also Kley 2000), and the most likely outcome is that one
planet will be ejected whereas the other \delete{one}\add{planet} will end
up on an
eccentric orbit. This could account for the observed eccentricities of the
extrasolar planets which are not orbiting close to their host star, i.e. which have
not migrated all the way to the star.

\end{itemize}

\section{Acknowledgements}
We wish to thank J.C.B. Papaloizou, R.P. Nelson, C. Terquem, J.D. Larwood, A.A. Christou \add{and
an anonymous referee}
for useful comments and criticism.
This work was partially supported (for F.M.) by the research
network ``Accretion onto black holes, compact stars and protostars''
funded by the European Commission under contract number ERBFMRX-CT98-0195,
\add{and additionally supported (for M.S.) by funding from a PPARC
research studentship.} 
Computational resources of the Grand HPC consortium were available and are gratefully 
acknowledged. We thank Udo Ziegler for making a FORTRAN version of his code NIRVANA
publicly available.


\begin{thebibliography}{}

%

\bibitem{}
Balbus S.A., Hawley J.F., 1991, ApJ, 376, 214 

\bibitem{}
Beaug\'e C., Aarseth S.J., Ferraz-Mello S., 1994, MNRAS, 270, 21 


\bibitem{}
Goldreich, P., Tremaine, S., 1979, ApJ, 233, 857

\bibitem{}
Haghighipour N., 1999, MNRAS, 304, 185 

\bibitem{}
Kley W., 1999, MNRAS, 303, 696 

\bibitem{}
Kley W., 2000, MNRAS, 313, 47



\bibitem{}
Malhotra, R., 1993, Icarus, 106, 264 

\bibitem{}
Marcy G.W., Cochran W.D., Mayor M., 1999, Protostars and Planets IV, Tucson: University of 
Arizona Press; eds Mannings, V., Boss, A.P., Russell, S., p. 1285. 

\bibitem{}
Masset F., 2000, A\&AS, 141, 165

\bibitem{}
Melita M.D., Woolfson M.M., 1996, MNRAS, 280, 854 

\bibitem{}
Nelson R.P., Papaloizou J.C.B., Masset F., Kley W., 2000, MNRAS, 318, 18


\bibitem{}
Papaloizou J.C.B., Lin D.N.C., 1984, ApJ, 285, 818 


\bibitem{}
Pollack, J.B., Hubickyj, O., Bodenheimer, P., Lissauer, J.J., Podolak, M., Greenzweig, Y., 1996, Icarus, 124, 62

\bibitem{}
Ward W.R, Hourigan K. 1989, ApJ, 347, 490 


\bibitem{}
Ward, W.R., 1991, Abstracts of the Lunar and Planetary Science Conference, 22, 1463 

\bibitem{}
Ward, W.R., 1992, Abstracts of the Lunar and Planetary Science Conference, 23, 1491 

\bibitem{}
Ward W.R., 1997, Icarus, 126, 261 

\bibitem{}
Ziegler U., Yorke H.W., 1997, Comp. Phys. Comp., 101, 54 

\end{thebibliography}
\end{document}